# Some Theoretical Results Concerning non-Parametric Estimation by Using a Judgment Post-stratification Sample

Ali Dastbaravarde[1], Nasser Reza Arghami, Majid Sarmad

*Department of Statistics, School of Mathematical Sciences, Ferdowsi University of Mashhad, P.O. Box 91775-1159, Mashhad, Iran*

**Abstract.** In this paper, some of the properties of non-parametric estimation of the expectation of *g(X)* (any function of *X*), by using a Judgment Post-stratification Sample (JPS), are discussed. A class of estimators (including the standard JPS estimator and a JPS estimator proposed by Frey and Feeman (2012, Comput. Stat. Data An.)) is considered. The paper provides mean and variance of the members of this class, and examines their consistency and asymptotic distribution. Specifically, the results are for the estimation of population mean, population variance and CDF. We show that any estimators of the class may be less efficient than Simple Random Sampling (SRS) estimator for small sample sizes. We prove that the relative efficiency of some estimators in the class with respect to Balanced Ranked Set Sampling (BRSS) estimator tends to 1 as the sample size goes to infinity. Furthermore, the standard JPS mean estimator and, Frey and Feeman JPS mean estimator are specifically studied and we show that two estimator have the same asymptotic distribution. For the standard JPS mean estimator, in perfect ranking situations, optimum values of $H$ (the ranking class size), for different sample sizes, are determined non-parametrically for populations that are not heavily skewed or thick tailed. We show that the standard JPS mean estimator may be more efficient than BRSS for large sample sizes, in situations in which we can use a larger class size for $H$ in JPS set-up.

---

[1] Corresponding author. E-mail addresses: ali.dastbaravarde@stu.um.ac.ir (A. Dastbaravarde).

# 1 Introduction

MacEachern et al. (2004) proposed Judgment Post-stratification Sampling (JPS) as an alternative to Ranked Set Sampling (RSS), introduced by McIntyre (1952). JPS is similar to RSS in that both designs use rank information to improve estimates. To obtain a sample of size $n$ under JPS design, one takes a Simple Random Sample (SRS), $X_{1,\ldots,}X_n$, and measures all the observations. Then, for each $X_i$, $i=1,\ldots,n$, one takes an auxiliary SRS $X_{i2},\ldots,X_{iH}$ of size $H-1$ from the same population, without actually measuring the units in the auxiliary samples. For each $i=1,\ldots,n$, the rank $R_i$ of $X_i$ in the $i^{\text{th}}$ ranking class=$\{X_i, X_{i2},\ldots,X_{iH}\}$ is determined by judgment. The JPS data, therefore, consists of $n$ i.i.d pairs $\{(X_i, R_i), i=1,\ldots,n\}$. The data set is called *full rank* if, for each $r=1,\ldots,H$, there exists at least one $R_i$ equal to $r$.

Judgment post-stratification sampling is similar in theory and application to ranked set sampling. Both methods are used in situations where full measurement of characteristic of interest is expensive while there is an inexpensive method by which observations can be ranked. In JPS and RSS, ranking information is used to artificially create a stratified sample that brings about more precise estimators than a simple random sample with the same sample size. In fact, JPS may be considered as a randomized version of Balanced RSS (BRSS).

The number of measured observations of each rank is typically fixed in RSS, while in JPS this is random. In JPS, it may be that there is not any measured observation for one or more ranks in the final sample and this is a restriction of JPS. Hence, JPS is usually less efficient than BRSS. But JPS is more flexible than RSS, since a researcher starts with an SRS, and then he/she may add rank information to improve the inference, if necessary. Moreover, in situations where ranking is imperfect, multiple rankers can be used or single ranker can be permitted to declare ties or both. In addition, JPS may allow for a greater size of ranking classes (i.e., $H$) in some applications.

MacEachern et al. (2004) showed that rankers can state uncertainty about ranks. They introduced estimators for mean and variance with imperfect ranking. They also showed by a simulation study, that multiple rankers can provide better estimates than a single ranker. Stokes et al. (2007) developed combining ranking information from multiple rankers for estimating the population mean. Wang et al. (2006) generalized definition of a concomitant of an order statistic in the multivariate case, and they applied this theory to develop some estimators of the mean. Wang et al. (2008) used a stochastic ordering assumption on in-stratum distributions to do more efficient inference. They proposed a mean estimator using isotonic regression. Frey and Ozturk (2011) modified the stochastic ordering constraint. They showed that the in-stratum cumulative distribution functions (CDFs) can be no more extreme than the CDFs for order statistics from the overall distribution, and they, in the JPS case, obtained better small-sample estimates of the overall CDF and the population mean. Ozturk (2012) provides a sampling scheme for JPS to combine the judgment ranks of rankers, which improves inference about mean and CDF of population. Wang et al. (2012) dealt with the empty strata in the proposed sampling models. They proposed modified isotonized estimators to improve estimation efficiency of CDF. Recent work by Frey and Feeman (2012) determined the optimum estimator in a class of unbiased linear estimators and showed that the standard JPS mean estimator is inadmissible.

In this paper, we expand the theory of JPS estimation by considering the problem of non-parametric estimation of the expectation of *g(X)* (any function of *X*) by using a JP sample. We consider a class of estimators and provide mean and variance of the members of this class and examine their consistency and asymptotic distribution in Section 2. We also specifically present the results for estimation of population mean, population variance and CDF. In section 3, we compare performance of the members of the class with SRS and BRSS estimators by examining their relative efficiency and asymptotic relative efficiency. We show that the member of the class (including the standard JPS estimator and Frey & Feeman JPS estimator) may be less efficient than SRS for small sample sizes for some distributions. Furthermore, we specifically examined the estimation of population mean. We discuss the performance of the standard JPS mean estimator, in comparison with the SRS, BRSS mean estimators and the JPS mean estimator proposed by Frey and Feeman (2012), in perfect ranking situation. Also, we show that the standard JPS mean estimator may be more efficient than BRSS for large sample sizes, in situations in which we can use a larger class size for *H* in JPS set-up. In section 4, we obtain the optimal *H* for the standard JPS mean estimator with perfect ranking for different sample sizes for populations that are not heavily skewed or thick tailed. A conclusion follows in section 5.

# 2 Estimation of mean of any function of a random variable using JPS data

Suppose $X$ is the variable of interest and a JPS sample of size $n$ is drawn from the population, giving us $n$ i.i.d pairs $(X_i, R_i)$, $i = 1,...,n$. Note that $R_i$, $i=1,\ldots,n$, is a discrete uniform random variable on the set of ranks 1 through $H$. We set $\mathbf{R} = (R_1,...,R_n)$. Define $I_{ir} = 1$ if observation $X_i$ has judgment rank $r$ ($R_i = r$), otherwise $I_{ir} = 0$, for every $i$ and $r$. We denote the number of the observations $X_i$, with judgment rank $r$, by $N_r$, thus $N_r = \sum_{i=1}^{n} I_{ir}$. Thus, $N_r$ is the number of the observations which fall into the post-stratum $r$. Note that $\mathbf{N} = (N_1,...,N_H)$ is a random vector with multinomial distribution with parameters $(n, 1/H,...,1/H)$. Define $I_r = 1$ if there is at least one measured observation in the $r^{\text{th}}$ post-stratum (i.e. $N_r > 0$), otherwise $I_r = 0$, for $r = 1,...,H$; and define $h_n = \sum_{r=1}^{H} I_r$ i.e. the number of observed post-strata in the sample. Also, define $J_r = 1/N_r$ if $N_r > 0$, otherwise $J_r = 0$, for $r = 1,...,H$.

Let $X_{[r]}$ denote $X \,|R = r$, any observation that fall in post-stratum $r$, $r=1,\ldots,H$. The brackets are used to indicate that rank of the observation is determined by judgment and may be in error. If the ranking is perfect (without error), the brackets are replaced with parentheses. In this case, $X_{(r)}$ is the $r^{th}$ order statistic in a simple random sample of size *H*. We can consider $X_{[r]}$'s, *r=1,…,H*, to be judgment order statistics.

Let $\mu_g$ denote the expectation of $g(X)$, where *g* is any function. We consider estimation of $\mu_g$ in the JPS set-up. For example, $g(x)=x^k$, $k=1,2,3,...$, corresponds to the estimation of population moments and $g(x)=I_{\{x\le c\}}$ corresponds to the estimation of CDF. We, as in Frey and Feeman (2012), consider the class $\mathcal{C}_{\mu_g}$ of estimators $\sum_{r=1}^{H} C_r \overline{g(X_{[r]})}$, where

$$\overline{g(X_{[r]})}=\begin{cases} 0 & N_r=0 \\ \frac{1}{N_r}\sum_{i=1}^{n} g(X_i) I_{ir} & N_r>0, \end{cases}$$

and the weights $C_r$ satisfy the following conditions

1. $\sum_{r=1}^{H} C_r = 1$.
2. $C_r = 0$ if $N_r = 0$, *r=1,…,H.*
3. The weights $C_r$, *r=1,…,H,* are random variables only through $\mathbf{N}$.
4. The weights $C_r$, *r=1,…,H,* depend on $N_1,...,N_H$ in a symmetric way.

Thus, $\mathcal{C}_{\mu_g}$ is a class of randomly weighted average of mean of the post-strata ($\overline{g(X_{[r]})}$) such that the weights of post-strata remain unchanged by permuting the post-strata. Note that the $C_r$'s are identically distributed.

**Remark 1.** In this approach, the class of population mean estimators and cumulative distribution function (CDF) estimators are given, respectively, by $\sum_{r=1}^{H} C_r \overline{X_{[r]}}$ and $\sum_{r=1}^{H} C_r \hat{F}_{[r]}(x)$, where $\overline{X_{[r]}}$ and $\hat{F}_{[r]}(x)$ are the mean and the empirical CDF of observations in the post-stratum *r*, respectively. Also, we can estimate population variance $\sigma^2$ by estimating $\mu = E(X)$ and $\mu_2 = E(X^2)$, that is $\hat{\sigma}^2_{JPS,C} = \sum_{r=1}^{H} C_r \overline{X^2_{[r]}} - \left(\sum_{r=1}^{H} C_r \overline{X_{[r]}}\right)^2$.

**Remark 2.** The class $\mathcal{C}_{\mu_g}$ includes SRS estimator ($C_r = \frac{N_r}{n}$)

$$\hat{\mu}_{g;SRS} = \frac{1}{n}\sum_{i=1}^{n} g(X_i),$$

standard JPS estimator ($C_r = \frac{I_r}{h_n}$)

$$\hat{\mu}_{g;JPS} = \frac{1}{h_n}\sum_{r=1}^{H} \overline{g\left(X_{[r]}\right)},$$

and Frey and Feeman estimator ($C_r = A_r$)

$$\tilde{\mu}_{g;JPS} = \sum_{r=1}^{H} A_r \overline{g\left(X_{[r]}\right)},$$

where $A_r = a_r \Big/ \sum_{s=1}^{H} a_s$ , where $a_r = \frac{N_r}{HN_r + 2}$.

### 2.1 Some properties

In this subsection, we study mean, variance, consistency and asymptotic distribution of the estimators of the class $\mathcal{C}_{\mu_g}$ . To study the properties, we need the following preliminary lemma.

**Lemma 1.** Let $\left\{X_{[1]},...,X_{[H]}\right\}$ be any set of judgment order statistics of i.i.d random variables $X_1,...,X_H$ and let g be any function. Also, suppose $g\left(X_1\right)$ has the mean $\mu_g$ and finite variance $\sigma_g^2$; and for any r, $\mu_g\left[r\right]$ and $\sigma_g^2\left[r\right]$ are the mean and the variance of $g\left(X_{[r]}\right)$, respectively. Then

(i) $$\mu_g = \frac{1}{H}\sum_{r=1}^{H}\mu_g\left[r\right],$$

(ii) $$\sigma_g^2 = \frac{1}{H}\sum_{r=1}^{H}\sigma_g^2\left[r\right] + \frac{1}{H}\sum_{r=1}^{H}\left(\mu_g\left[r\right] - \mu_g\right)^2.$$

***Proof:*** (i) By using the iterative expectation, we can write

$$\begin{aligned} E\left(g\left(X_1\right)\right) &= E\left(E\left[g\left(X_1\right) \middle| R_1\right]\right) \\ &= \frac{1}{H}\sum_{r=1}^{H} E\left[g\left(X_1\right) \middle| R_1 = r\right] \\ &= \frac{1}{H}\sum_{r=1}^{H} E\left[g\left(X_{[r]}\right)\right]. \end{aligned}$$

(ii) Using the conditional variance formula, we can write

$$
\begin{aligned}
V\left(g\left(X_1\right)\right) &= E\left(V\left[g\left(X_1\right)|R_1\right]\right)+V\left(E\left[g\left(X_1\right)|R_1\right]\right) \\
&= \frac{1}{H}\sum_{r=1}^{H} V\left[g\left(X_1\right)|R_1=r\right]+\frac{1}{H}\sum_{r=1}^{H}\left(E\left[g\left(X_1\right)|R_1=r\right]-E\left[g\left(X_1\right)\right]\right)^2 \\
&= \frac{1}{H}\sum_{r=1}^{H} V\left[g\left(X_{[r]}\right)\right]+\frac{1}{H}\sum_{r=1}^{H}\left(E\left[g\left(X_{[r]}\right)\right]-E\left[g\left(X_1\right)\right]\right)^2 . \square
\end{aligned}
$$

The above lemma, where $g$ is the identity function is proved by Dell and Clutter (1972). Lemma 1 provides that the total population mean is equal to the mean of the post-stratum means, and the total population variance is equal to the mean of variances of within post-strata ($\frac{1}{H}\sum_{r=1}^{H}\sigma_g^2[r]$) plus the variance of between post-strata ($\frac{1}{H}\sum_{r=1}^{H}\left(\mu_g[r]-\mu_g\right)^2$). Now, the following theorem gives the mean and variance of the estimators of the class $\mathcal{C}_{\mu_g}$ .

***Theorem 1.*** *Let $\left(X_i, R_i\right)$, i=1,…,n, be a JPS sample and $g\left(X_i\right)$ have finite variance $\sigma_g^2$. Then,*

*(i) All estimators in the class $\mathcal{C}_{\mu_g}$ are unbiased for $\mu_g$.*

*(ii) The variance for an estimator in the class $\mathcal{C}_{\mu_g}$ is given by*

$$
E\left(J_1 C_1^2\right)\sum_{r=1}^{H}\sigma_g^2[r]+\frac{H}{H-1}V\left(C_1\right)\left(\sum_{r=1}^{H}\left(\mu_g[r]-\mu_g\right)^2\right).
$$

***Proof:*** (i) We can write

$$
\begin{aligned}
E\left(\sum_{r=1}^{H} C_r \overline{g\left(X_{[r]}\right)}\right) &= E\left(\sum_{r=1}^{H} C_r E\left[\overline{g\left(X_{[r]}\right)}|\mathbf{R}\right]\right) \\
&= E\left(\sum_{r=1}^{H} C_r \mu_g[r]\right) \\
&= E\left(C_1\right)\sum_{r=1}^{H}\mu_g[r] \\
&= \frac{1}{H}\sum_{r=1}^{H}\mu_g[r]=\mu_g .
\end{aligned} \tag{1}
$$

The first equality holds because $C_r$ ’s are functions of $\mathbf{N}$. The second equality holds because

$$
E\left(\overline{g\left(X_{[r]}\right)}|\mathbf{R}=\mathbf{r}\right)=\begin{cases} 0 & n_r=0 \\ E\left(\frac{1}{N_r}\sum_{i=1}^{n} g\left(X_i\right) I_{ir}|\mathbf{R}=\mathbf{r}\right) & n_r>0, \end{cases}
$$

and

$$E\left(\frac{1}{N_r}\sum_{i=1}^{n} g\left(X_i\right) I_{ir} \middle| \mathbf{R}=\mathbf{r}\right)=\frac{1}{n_r}\sum_{i=1}^{n} E\left(g\left(X_i\right)\middle| R_i=r\right) I_{ir}$$

$$=\frac{1}{n_r}\sum_{i=1}^{n}\mu_g[r] I_{ir}=\mu_g[r].$$

The third equality in (1) holds because $C_r$ 's are identically distributed. The equality before the last in (1) holds because we have (as proved in the Appendix A) $E\left(C_1\right)=1/H$ . And the last equality in (1) holds because of part (i) of Lemma 1.

(ii) We can write

$$V\left(\sum_{r=1}^{H} C_r \overline{g\left(X_{[r]}\right)}\right)=V\left[E\left(\sum_{r=1}^{H} C_r \overline{g\left(X_{[r]}\right)}\middle|\mathbf{R}\right)\right]+E\left[V\left(\sum_{r=1}^{H} C_r \overline{g\left(X_{[r]}\right)}\middle|\mathbf{R}\right)\right],$$

where

$$V\left[E\left(\sum_{r=1}^{H} C_r \overline{g\left(X_{[r]}\right)}\middle|\mathbf{R}\right)\right]=V\left[\sum_r C_r \mu_g[r]\right]$$

$$=\sum_r \mu_g^2[r]\, V\left(C_r\right)+2\sum_{r<s}\mu_g[r]\mu_g[s] COV\left(C_r, C_s\right)$$

$$=V\left(C_1\right)\sum_r \mu_g^2[r]+2 COV\left(C_1, C_2\right)\sum_{r<s}\mu_g[r]\mu_g[s]$$

$$=V\left(C_1\right)\sum_r \mu_g^2[r]-\frac{1}{H-1} V\left(C_1\right)\left[H^2\mu_g^2-\sum_r \mu_g^2[r]\right]$$

$$=\frac{H}{H-1} V\left(C_1\right)\left(\sum_{r=1}^{H}\left(\mu_g[r]-\mu_g\right)^2\right).$$

The third equality holds by the fact that $C_r$ 's are identically distributed. The equality before the last holds because we have $COV\left(C_1, C_2\right)=\frac{-1}{H-1} V\left(C_1\right)$ (as proved in the Appendix A) and $H^2\mu_g^2=\sum_{r=1}^{H}\mu_g^2[r]+2\sum_{r<s}\mu_g[r]\mu_g[s]$.

We can also write

$$E\left[V\left(\sum_{r=1}^{H} C_r \overline{g\left(X_{[r]}\right)}\middle|\mathbf{R}\right)\right]=E\left[\sum_r C_r^2 V\left(\overline{g\left(X_{[r]}\right)}\middle|\mathbf{R}\right)\right]$$

$$=E\left[\sum_r J_r C_r^2 \sigma_g^2[r]\right]$$

$$=E\left(J_1 C_1^2\right)\sum_r \sigma_g^2[r].$$

The first equality holds because, given $\mathbf{R}$, $\overline{g\left(X_{[r]}\right)}$'s are conditionally independent. The second equality holds because

$$V\left(\overline{g\left(X_{[r]}\right)}|\mathbf{R}=\mathbf{r}\right)=\begin{cases}0 & n_r=0\\ \sigma_g^2[r]/n_r & n_r>0\end{cases}. \square$$

**Remark 3.** By using part (ii), the variance for an estimator in $\mathcal{C}_{\mu_g}$ can also be written as

$$H\,E\left(J_1C_1^2\right)\sigma_g^2-\left[H\,E\left(J_1C_1^2\right)-\frac{H^2}{H-1}V\left(C_1\right)\right]\left(\frac{1}{H}\sum_{r=1}^{H}\left(\mu_g[r]-\mu_g\right)^2\right).$$

**Corollary 1.** Let $(X_i,R_i)$, $i=1,\ldots,n$, be a JPS sample.

(i) if $X$ has finite variance $\sigma^2$ then, $\hat{\mu}_{JPS}=\frac{1}{h_n}\sum_{r=1}^{H}\overline{X_{[r]}}$ and $\tilde{\mu}_{JPS}=\sum_{r=1}^{H}A_r\overline{X_{[r]}}$ are two unbiased estimator for population mean, $\mu$, with respective variances

$$V\left(\hat{\mu}_{JPS}\right)=E\left(\frac{J_1}{h_n^2}\right)\sum_{r=1}^{H}\sigma_{[r]}^2+\frac{H}{H-1}V\left(\frac{I_1}{h_n}\right)\left(\sum_{r=1}^{H}\left(\mu_{[r]}-\mu\right)^2\right),$$

and

$$V\left(\tilde{\mu}_{JPS}\right)=E\left(J_1A_1^2\right)\sum_{r=1}^{H}\sigma_{[r]}^2+\frac{H}{H-1}V\left(A_1\right)\left(\sum_{r=1}^{H}\left(\mu_{[r]}-\mu\right)^2\right),$$

where $\mu_{[r]}$ and $\sigma_{[r]}^2$ are the mean and the variance of $r^{th}$ judgment order statistics, $X_{[r]}$, respectively.

(ii) For any $x$, an estimator $\sum_{r=1}^{H}C_r\hat{F}_{[r]}(x)$ is unbiased for CDF, $F(x)$, and its variance is given by

$$H\,E\left(J_1C_1^2\right)\left[F(x)\left(1-F(x)\right)\right]-\left[E\left(J_1C_1^2\right)-\frac{H}{H-1}V\left(C_1\right)\right]\left(\sum_{r=1}^{H}\left(F_{[r]}(x)-F(x)\right)^2\right).$$

The above corollary shows that the variance of the estimators $\hat{\mu}_{JPS}$ and $\tilde{\mu}_{JPS}$ are linear combinations of mean of the variances of within post-strata and variance of between post-strata. Also, the coefficients in these linear combinations depend on $n$ and $H$ only and not the distribution of $X$. Coefficients $E\left(J_1/h_n^2\right)$ and $V\left(I_1/h_n\right)$ in the variance of $\hat{\mu}_{JPS}$ were computed analytically and are given in Appendix B.

**Remark 4.** Frey and Feeman (2012), using a different method, showed that all estimator in the class $\sum_{r=1}^{H} C_r \overline{X_{[r]}}$ are unbiased. They also provided the conditional variance, given the ordered sample sizes of post-strata, for any estimator in class $\sum_{r=1}^{H} C_r \overline{X_{[r]}}$.

**Corollary 2.** For a set $\{C_1,...,C_H\}$ such that for any $r$, $C_r \neq N_r/n$, the estimator $\sum_{r=1}^{H} C_r \overline{g\left(X_{[r]}\right)}$ is dominated by SRS estimator if and only if $V\left(C_1\right) \geq V\left(N_1/n\right)$.

**Proof:** We can write

$$V\left(\sum_{r=1}^{H} C_r \overline{g\left(X_{[r]}\right)}\right) - V\left(\frac{1}{n}\sum_{i=1}^{n} g\left(X_i\right)\right) = \left[HE\left(J_1 C_1^2\right) - \frac{1}{n}\right]\frac{1}{H}\sum_{r=1}^{H}\sigma_g^2[r]$$
$$+\left[\frac{H^2}{H-1}V\left(C_1\right) - \frac{1}{n}\right]\frac{1}{H}\sum_{r=1}^{H}\left(\mu_g[r]-\mu_g\right)^2.$$

Note that, for any $C_1 \neq N_1/n$, $nHE\left(J_1 C_1^2\right) > 1$, as proved in Appendix A. Therefore, $V\left(\sum_{r=1}^{H} C_r \overline{g\left(X_{[r]}\right)}\right) > V\left(\sum_{i=1}^{n} g\left(X_i\right)/n\right)$ if and only if $V\left(C_1\right) \geq (H-1)/\left(nH^2\right)$.

**Remark 5.** Note that if $C_r = I_r/h_n$ or $A_r$ then, for $n \geq 3$ and $H \geq 2$, $V\left(C_r\right) < \frac{H-1}{nH^2}$, as proved in Appendices B and C.

The following theorem establishes asymptotic properties of the estimators of the class $\mathcal{C}_{\mu_g}$.

***Theorem 2.*** *Let the conditions of Theorem 1 hold. Then, for any set of weights $\{C_1,...,C_H\}$ that*

$$\sqrt{n}\left(C_r - \frac{1}{H}\right) \xrightarrow{P} 0 \quad \forall r \tag{2}$$

*as $n \to \infty$,*

*(i) The estimator $\sum_{r=1}^{H} C_r \overline{g\left(X_{[r]}\right)}$ is consistent.*

*(ii) For fixed $H$, as $n \to \infty$*

$$\sqrt{n}\left(\sum_{r=1}^{H} C_r \overline{g\left(X_{[r]}\right)} - \mu_g\right) \xrightarrow{d} N\left(0\,,\ \frac{1}{H}\sum_{r=1}^{H}\sigma_g^2[r]\right).$$

***Proof:*** (i) From (2), we, for any *r*, have that $C_r$ converges in probability to $1/H$ as $n \to \infty$. Also, for any *r*, $nJ_r$ and $\frac{1}{n}\sum_{i=1}^{n} g(X_i)I_{ir}$ converge almost surely (a.s.) to *H* and $\frac{1}{H}\mu_g[r]$, respectively. So, $\overline{g(X_{[r]})}$ converges a.s. to $\mu_g[r]$ as $n \to \infty$. And this proves the result.

(ii) We can write

$$\sqrt{n}\left(\sum_{r=1}^{H} C_r \overline{g(X_{[r]})} - \mu_g\right) = \sum_{r=1}^{H} \sqrt{n} C_r \left(\overline{g(X_{[r]})} - \mu_g[r]\right) + \sum_{r=1}^{H}\left[\mu_g[r]\sqrt{n}\left(C_r - \frac{1}{H}\right)\right].$$

We set $\mathbf{T}_i = \left((g(X_i) - \mu_g[1])I_{i1}, \ldots, (g(X_i) - \mu_g[H])I_{iH}\right)^T$, *i=1,…,n*. Hence, $\mathbf{T}_i$ 's are i.i.d random vectors with mean $\mathbf{0}$ and variance-covariance matrix $\mathbf{\Sigma} = [\sigma_{rs}]_{H \times H}$ where

$$\begin{aligned}\sigma_{rr} &= V\left[(g(X_i) - \mu_g[r])I_{ir}\right] = \frac{1}{H}\sigma_g^2[r] \quad \forall r, \\ \sigma_{rs} &= COV\left[(g(X_i) - \mu_g[r])I_{ir}, (g(X_i) - \mu_g[s])I_{is}\right] = 0 \quad \forall r \neq s.\end{aligned}$$

Using central limit theorem

$$\frac{\sum_{i=1}^{n} \mathbf{T}_i}{\sqrt{n}} = \frac{1}{\sqrt{n}}\left(N_1\left(\overline{g(X_{[1]})} - \mu_g[1]\right), \ldots, N_H\left(\overline{g(X_{[H]})} - \mu_g[H]\right)\right)^T \xrightarrow{d} N(\mathbf{0}, \mathbf{\Sigma}).$$

Besides

$$(nJ_1C_1, \ldots, nJ_HC_H)^T \xrightarrow{P} (1, \ldots, 1)^T.$$

So by Slutsky theorem

$$\sqrt{n}\sum_{r=1}^{H} C_r\left(\overline{g(X_{[r]})} - \mu_g[r]\right) = \frac{1}{\sqrt{n}}(nJ_1C_1, \ldots, nJ_HC_H)\sum_{i=1}^{n}\mathbf{T}_i \xrightarrow{d} N\left(0, \frac{1}{H}\sum_{r=1}^{H}\sigma_g^2[r]\right).$$

Furthermore, we have $\sum_{r=1}^{H}\left[\mu_g[r]\sqrt{n}(C_r - 1/H)\right]$ converges in probability to 0 as $n \to \infty$. This completes the proof.□

**Corollary 3.** Let $(X_i, R_i)$, $i = 1, \ldots, n$, be a JPS sample and $X$ have finite variance $\sigma^2$. Then

(i) The estimators $\hat{\mu}_{JPS}$ and $\tilde{\mu}_{JPS}$ are strongly consistent for $\mu$.

(ii) For fixed *H*, $\sqrt{n}(\hat{\mu}_{JPS} - \mu)$ and $\sqrt{n}(\tilde{\mu}_{JPS} - \mu)$ converge in distribution to the same distribution, $N\left(0, \frac{1}{H}\sum_{r=1}^{H}\sigma_{[r]}^2\right)$ as $n \to \infty$.

**Proof:** (i) For each *r*, $I_r$ is Bernoulli random variable with parameter $1-\left(\frac{H-1}{H}\right)^n$. So, $I_r$ converges a.s. to 1 as $n\to\infty$. Hence, $h_n$ converges a.s. to *H* as $n\to\infty$. Furthermore, $N_r/n$ converges a.s. to $1/H$ as $n\to\infty$. Hence, for any *r*, $A_r$ converges a.s. to $1/H$ as $n\to\infty$. From part (i) Theorem 2, $\bar{X}_{[r]}$ converges a.s. to $\mu_{[r]}$ as $n\to\infty$. And this completes the proof.

(ii) We have (as proved in Appendices B and C) $\sqrt{n}\left(\frac{1}{h_n}-\frac{1}{H}\right)$ and $\sqrt{n}\left(A_r-\frac{1}{H}\right)$ converges a.s. to 0 as $n\to\infty$. So, part (ii) Theorem completes the proof. □

**Remark 6.** Using Theorem 1, it is easy to show that $E\left(\hat{\sigma}^2_{JPS,C}\right)=\sigma^2-V\left(\sum_{r=1}^{H} C_r \overline{X_{[r]}}\right)$, that is $\hat{\sigma}^2_{JPS,C}$ underestimates $\sigma^2$. But $\hat{\sigma}^2_{JPS,C}$ is asymptotically unbiased for $\sigma^2$ and, in addition, $\hat{\sigma}^2_{JPS,C}$ is strongly consistent by Theorem 2. Also, see McEachern et al. (2002) for an unbiased and efficient estimator of the population variance.

**Remark 7.** Let $Y=g(X)$ and suppose *Y* observations can be ranked perfectly and with negligible cost. Under such conditions, mean of $g(X)$ can also be estimated by ranking $Y_i$'s, $i=1,\ldots,n$, in their ranking classes, and using the formula $\sum_{r=1}^{H} C_r \overline{Y_{\{r\}}}$. The notation $\{\ \}$ is used to indicate that precision of judgment ranking of *Y* observation may be different. The properties of this estimator trivially follow our results about $\sum_{r=1}^{H} C_r \overline{X_{[r]}}$. That is $\sum_{r=1}^{H} C_r \overline{Y_{\{r\}}}$ is unbiased for $\mu_g$ and its variance is

$$V\left(\sum_{r=1}^{H} C_r \overline{Y_{\{r\}}}\right)=K_1\frac{1}{H}\sum_{r=1}^{H} V\left(Y_{\{r\}}\right)+K_2\left(\frac{1}{H}\sum_{r=1}^{H}\left(E\left(Y_{\{r\}}\right)-\mu_g\right)^2\right),$$

where $K_1=H\,E\left(J_1C_1^2\right)$ and $K_2=\frac{H^2}{H-1}V\left(C_1\right)$. Hence, we have

$$V\left(\sum_{r=1}^{H} C_r \overline{Y_{\{r\}}}\right)-V\left(\sum_{r=1}^{H} C_r \overline{g\left(X_{[r]}\right)}\right)=\frac{K_1-K_2}{H}\left\{\sum_{r=1}^{H}\left[E\left(g\left(X_{[r]}\right)\right)\right]^2-\sum_{r=1}^{H}\left[E\left(g\left(X\right)_{\{r\}}\right)\right]^2\right\},$$

According to Corollary 2, for $C_r$'s that $V\left(C_r\right)<\frac{H-1}{nH^2}$, $K_1-K_2>0$; therefore, variance of $\sum_{r=1}^{H} C_r \overline{Y_{\{r\}}}$ is less than variance of $\sum_{r=1}^{H} C_r \overline{g\left(X_{[r]}\right)}$ if ranking *X* observations are not more precise than ranking *Y* observations. In the situation that precision of ranking *X* observations and *Y* observations are equal if *g* is one-one then it is intuitively clear and easy to prove that two variances $\sum_{r=1}^{H} C_r \overline{Y_{\{r\}}}$ and $\sum_{r=1}^{H} C_r \overline{g\left(X_{[r]}\right)}$ are equal; otherwise, variance of $\sum_{r=1}^{H} C_r \overline{Y_{\{r\}}}$ is less than variance of $\sum_{r=1}^{H} C_r \overline{g\left(X_{[r]}\right)}$. In situations that ranking $X$ and $Y$ are perfect, if $g$ is not one-one, then variance of $\sum_{r=1}^{H} C_r \overline{Y_{(r)}}$ is less than variance of $\sum_{r=1}^{H} C_r \overline{g\left(X_{(r)}\right)}$. This follows from the following lemma.

***lemma 2.*** *Suppose $S=\{X_1,\ldots,X_k\}$ is a set of random variables, not necessarily independent or identically distributed. Then (assuming the moments exist),*

$$\sum_{j=1}^{k}\left(\mu_{(j)}-\bar{\mu}\right)^2 \geq \sum_{j=1}^{k}\left(\mu_j-\bar{\mu}\right)^2, \tag{3}$$

*where* $\bar{\mu}=\frac{1}{k}\sum_{j=1}^{k}\mu_j=\frac{1}{k}\sum_{j=1}^{k}\mu_{(j)}$, $\mu_j=E\left(X_j\right)$, $\mu_{(j)}=E\left(X_{(j)}\right)$, $j=1,\ldots,k$ , *where* $X_{(j)}$ *is the* $j^{th}$ *element in ordered* $S$ .

***Proof.*** Let $\left(x_{i1},\ldots,x_{ik}\right)$, $i=1,\ldots,n$ be $n$ i.i.d observations of the random vector $\mathbf{X}=\left(X_1,\ldots,X_k\right)$. Then, by Lemma 8 (as proved in the Appendix D), we have

$$\sum_{j=1}^{k}\left(\frac{1}{n}\sum_{i=1}^{n}x_{i(j)}\right)^2 \geq \sum_{j=1}^{k}\left(\frac{1}{n}\sum_{i=1}^{n}x_{ij}\right)^2.$$

It follows from Strong Law of Large numbers that

$$\sum_{j=1}^{k}\mu_{(j)}^2 \geq \sum_{j=1}^{k}\mu_j^2. \tag{4}$$

Now it is elementary that (4) is equivalent to (3). Thus the lemma is proved.□

In practice, ranking $Y$ observations with negligible cost may not be possible or is less precise than ranking $X$ observations. Under such circumstance, which is often the case, $\sum_{r=1}^{H} C_r \overline{g\left(X_{[r]}\right)}$ may be the only practical estimator to use.

# 3 Comparisons

In this section, we compare the performance of the JPS estimators with respect to the corresponding estimators in SRS and BRSS set-up in general case, estimation of $\mu_g$, and in special case, estimation of population mean in perfect ranking situation.

### 3.1 General case

Using Theorem 1, the Relative Efficiency (RE) of $\sum_{r=1}^{H} C_r \overline{g\left(X_{[r]}\right)}$ with respect to $\hat{\mu}_{g;SRS}$ is

$$\mathrm{RE}\left(\sum_{r=1}^{H} C_r \overline{g\left(X_{[r]}\right)}, \hat{\mu}_{g;SRS}\right) = \left[M_1\left(1-\delta_g\right) + M_2\delta_g\right]^{-1},$$

where $M_1 = nK_1$, $M_2 = nK_2$ and $\delta_g = \frac{1}{H\sigma_g^2}\sum_{r=1}^{H}\left(\mu_g[r] - \mu_g\right)^2$.

Note that $\mathrm{RE}\left(\sum_{r=1}^{H} C_r \overline{g\left(X_{[r]}\right)}, \hat{\mu}_{g;SRS}\right)$ is location and scale invariant and depends on the distribution of $X$ only through $\delta_g$. Also note that $\sum_{r=1}^{H} C_r \overline{g\left(X_{[r]}\right)}$ may be less efficient than $\hat{\mu}_{g;SRS}$; in fact, since, for $C_1 \neq N_1/n$, $nHE\left(J_1 C_1^2\right) > 1$ (as proved in the Appendix A), for populations with very low $\delta_g$, it may be that $\mathrm{RE}\left(\sum_{r=1}^{H} C_r \overline{g\left(X_{[r]}\right)}, \hat{\mu}_{g;SRS}\right) < 1$. The following corollary provides Asymptotic Relative Efficiency (ARE) of $\sum_{r=1}^{H} C_r \overline{g\left(X_{[r]}\right)}$ with respect to $\hat{\mu}_{g;SRS}$.

***Corollary 4.*** *If $\sqrt{n}\left(C_1 - 1/H\right)$ converges in probability to 0, as $n \to \infty$ then, asymptotic relative efficiency of $\sum_{r=1}^{H} C_r \overline{g\left(X_{[r]}\right)}$ with respect to $\hat{\mu}_{g;SRS}$ for fixed H, as $n \to \infty$ is*

$$\mathrm{ARE}\left(\sum_{r=1}^{H} C_r \overline{g\left(X_{[r]}\right)}, \hat{\mu}_{g;SRS}\right) = \left(1-\delta_g\right)^{-1}.$$

***Proof:*** Note that $\delta_g$ does not depend on the sample size. It can be seen from the proof of part (i) of Theorem 2 that $M_1$ and $M_2$ tend to 1 and 0, respectively, as sample size goes to infinity. Thus the corollary is proved.□

Since $\delta_g < 1$, at least for large sample sizes, $\sum_{r=1}^{H} C_r \overline{g\left(X_{[r]}\right)}$ is more efficient than $\hat{\mu}_{g;SRS}$.

***Corollary 5:*** *For fixed H,*

$$RE\left(\frac{1}{h_n}\sum_{r=1}^{H}\overline{g\left(X_{[r]}\right)}, \sum_{r=1}^{H} A_r \overline{g\left(X_{[r]}\right)}\right) \xrightarrow{n\to\infty} 1.$$

***Proof:*** We can write

$$RE\left(\frac{1}{h_n}\sum_{r=1}^{H}\overline{g\left(X_{[r]}\right)},\sum_{r=1}^{H}A_r\overline{g\left(X_{[r]}\right)}\right)=\frac{RE\left(\frac{1}{h_n}\sum_{r=1}^{H}\overline{g\left(X_{[r]}\right)},\hat{\mu}_{SRS}\right)}{RE\left(\sum_{r=1}^{H}A_r\overline{g\left(X_{[r]}\right)},\hat{\mu}_{SRS}\right)}.$$

So, Corollary 4 completes proof.□

Now, we compare the performances of the JPS and BRSS $\mu_g$-estimators when the sample sizes and the ranking class sizes in the two methods of sampling are equal. In BRSS with *m cycle* and ranking class size *H*, the sample size (the number of measured units) is equal to *mH*. Thus we take the ranking class size and sample size for both sampling methods to be *H* and *mH* respectively. The BRSS $\mu_g$-estimator and its variance are

$$\hat{\mu}_{g;BRSS}=\frac{1}{mH}\sum_{r=1}^{H}\sum_{j=1}^{m}g\left(X_{[r]j}\right)$$

and

$$V\left(\hat{\mu}_{g;BRSS}\right)=\frac{1}{mH^2}\sum_{r=1}^{H}\sigma_g^2[r],$$

respectively (Chen et. al., 2004). Relative Efficiency of $\sum_{r=1}^{H}C_r\overline{g\left(X_{[r]}\right)}$ with respect to $\hat{\mu}_{g;BRSS}$ is thus equal to

$$\mathrm{RE}\left(\sum_{r=1}^{H}C_r\overline{g\left(X_{[r]}\right)},\hat{\mu}_{g;BRSS}\right)=\frac{1-\delta_g}{M_1\left(1-\delta_g\right)+M_2\delta_g}.$$

Here, also, RE is location and scale invariant. Note that $\mathrm{RE}\left(\sum_{r=1}^{H}C_r\overline{g\left(X_{[r]}\right)},\hat{\mu}_{g;BRSS}\right)<1$, because

$$\frac{1-\delta_g}{M_1\left(1-\delta_g\right)+M_2\delta_g}<\frac{1-\delta_g}{M_1\left(1-\delta_g\right)}=\frac{1}{M_1}.$$

The following corollary provides ARE of $\sum_{r=1}^{H}C_r\overline{g\left(X_{[r]}\right)}$ with respect to $\hat{\mu}_{g;BRSS}$.

***Corollary 6.*** *If $\sqrt{n}\left(C_1-1/H\right)$ converges in probability to 0, as $n\to\infty$ then, asymptotic relative efficiency of $\sum_{r=1}^{H}C_r\overline{g\left(X_{[r]}\right)}$ with respect to $\hat{\mu}_{g;BRSS}$ for fixed H, as $n\to\infty$ is equal to one.*

Proof is similar to the proof of Corollary 4. □

Thus, for large sample sizes, JPS and BRSS $\mu_g$ -estimators have the same performance.

### 3.2 Special case

In this subsection, we try to examine specifically the performance of the standard JPS mean estimator, in comparison with the SRS, BRSS mean estimators and the JPS mean estimator proposed by Frey and Feeman (2012), in perfect ranking situation, and identify effects of sample size, ranking class size, and population distribution.

We use the results of the previous section to draw the relative efficiency plots of $\hat{\mu}_{JPS}$ with respect to $\hat{\mu}_{SRS}$ in Figures 1 and 2. Figure 1 plots the relative efficiency as a function of sample size for two ranking class sizes $H$=2 and 5, for a set of distributions containing normal, student's t-distribution with 3 degrees freedom, uniform, beta(0.5,0.5), exponential, chi-square(5), Pareto with shape parameters 2.5 and 4, and Weibull with shape parameters 0.5 and 1.5 (a set which includes symmetric, heavy tail, uniform, U-shaped, and skewed distributions), denoted by N, t(3), U, B(0.5,0.5), E, C(5), P(2.5), P(4), W(0.5), and W(1.5), respectively. Figure 1 shows that the standard JPS mean estimator is more efficient than SRS in most situations (with the exception of Pareto and Weibull distributions when the sample size is small). It also shows that skewness and kurtosis have a negative effect on the relative efficiency of $\hat{\mu}_{JPS}$ with respect to $\hat{\mu}_{SRS}$. In gamma and Pareto family of distributions, RE increases as shape parameters increases. In a family of t-distributions, RE increases as df increases. Moreover, Figure 1 shows that RE increases with sample size (except for Pareto and Weibull distributions for small sample sizes).

Figure 2 shows RE as a function of ranking class size $H$ for some distributions, for two sample sizes $n$=10 and 30. It is important to note RE is not necessarily an increasing function of $H$. At the beginning, RE increases with $H$ and then decreases as $H$ increases (except for Pareto and Weibull distributions). Moreover, it can be shown that for fixed $n$, as $H \to \infty$, both $\delta_I$ (where $I$ is the identity function) and $C_2$ tend to 1 and hence RE tends to 1. The reason is that, for fixed $n$, the probability that the data set is not full rank, and thus the frequency of empty strata, increases as $H$ increases. In contrast, as $H$ increase, rank of $X_i$ would be a better indicator of how large or small $X_i$ is. Hence, for fixed $n$, RE is an increasing function of $H$ for small values of $H$, and becomes a decreasing function of $H$ for large values of $H$.

From both Figures 1 and 2, it can be seen that for small values of $\delta_I$, RE can be less than 1. It is easy to see that the minimum value of $\delta_I$ for which $\text{RE}\left(\hat{\mu}_{JPS}, \hat{\mu}_{SRS}\right) \geq 1$ is equal to $\frac{C_1 - 1}{C_1 - C_2}$.

Figure 3 plots the above minimum value of $\delta_I$ as a function of sample size for two class sizes $H$=2 and 5. It shows that the minimum value of $\delta_I$ decreases with sample size, such that it converges to 0 as $n \to \infty$. Figure 3 also shows that the minimum value of $\delta_I$ increases with $H$. But, this is not necessarily a serious problem, because for all distributions, $\delta$ itself increases with $H$, too. Numerical calculations show that, for fixed $n(\geq 3)$, as $H \to \infty$, the minimum value of $\delta_I$ converges to a number less than 0.43; while, $\delta_I$ converges to 1 for all distributions.

We also note that RE increases with $\delta_I$ and it attains its maximum value when the population distribution is uniform. Takahasi and Wakimoto (1968) proved that $\delta_I \leq \frac{H-1}{H+1}$, and that equality holds only when the population distribution is uniform. Hence, the maximum value of ARE is equal to $(H+1)/2$.

Now, we examine the relative efficiency of the standard JPS mean estimator with respect to the BRSS mean estimator. We can write $\mathrm{RE}(\hat{\mu}_{JPS}, \hat{\mu}_{BRSS}) = (1-\delta_I)\mathrm{RE}(\hat{\mu}_{JPS}, \hat{\mu}_{SRS})$. Since, $\delta_I$ does not depend on $n$, the effects of sample size on $\mathrm{RE}(\hat{\mu}_{JPS}, \hat{\mu}_{BRSS})$ and $\mathrm{RE}(\hat{\mu}_{JPS}, \hat{\mu}_{SRS})$ are the same. Also, we can write $\mathrm{RE}(\hat{\mu}_{JPS}, \hat{\mu}_{BRSS}) = \left[M_1 + M_2\,\delta_I/(1-\delta_I)\right]^{-1}$. So, $\mathrm{RE}(\hat{\mu}_{JPS}, \hat{\mu}_{BRSS})$ decreases as $\delta_I$ increases.

The results are summarized in Figures 4 and 5. Figure 4 plots the relative efficiency as a function of sample size for two ranking class sizes $H$=2 and 5 for some distributions. Figure 4 shows that the standard JPS mean estimator is less efficient than the BRSS, in all situations. Also, Figure 4 shows that RE generally increases as sample size increases (except for small sample sizes). Figure 5 plots RE as a function of ranking class size for some distributions for two sample sizes n=10 and 30. Figure 5 shows that RE is a decreasing function of $H$, because the probability of having a full rank JPS sample decreases as $H$ increases.

In RSS, we need to determine the rank of all units within each ranking class, whereas in JPS, we only need the rank of each fully measured unit in its ranking class. Thus contrary to RSS, JPS may allow for a greater size of ranking class, $H$, in some applications. Hence, we can compare the performance of the JPS and BRSS mean estimator more comprehensively when the JPS ranking class size ($H_J$) is allowed to be greater than the BRSS ranking class size ($H_B$). Note that $\mathrm{RE}(\hat{\mu}_{JPS}, \hat{\mu}_{BRSS})$ decreases with $H_B$ and it is not necessarily an increasing function of $H_J$. At the beginning, RE increases with $H_J$ and then decreases as $H_J$ increases; and RE is maximized when $H_J$ is equal to $H_{opt}$ (See section 4 for $H_{opt}$).

RE is calculated for two BRSS usual ranking class sizes ($H_B$=3,5) and some JPS ranking class sizes ($H_J$=3,4,5,6,7,8,10,12, and 14), for some distributions, and two sample sizes ($n$=15,60). The results are presented in Table 1 and Table 2.

Table 1 shows that, in small sample sizes, increasing $H_J$ does not have a considerable effect on RE. But in large sample sizes, RE will be greater than 1 even when $H_J=H_B+1$ (Table 2).

Table 1. RE for two $H_B$, some $H_J$ for some distributions in $n=15$

| | $H_B$ | $H_J=3$ | $H_J=4$ | $H_J=5$ | $H_J=6$ | $H_J=7$ | $H_J=8$ | $H_J=10$ | $H_J=12$ | $H_J=14$ |
|---|---|---|---|---|---|---|---|---|---|---|
| **Normal** | **3** | 0.82 | 0.88 | 0.90 | 0.90 | 0.89 | 0.87 | 0.83 | 0.79 | 0.76 |
| | **5** | 0.57 | 0.61 | 0.62 | 0.62 | 0.61 | 0.60 | 0.57 | 0.54 | 0.52 |
| **t(3)** | **3** | 0.83 | 0.84 | 0.85 | 0.85 | 0.85 | 0.85 | 0.84 | 0.82 | 0.81 |
| | **5** | 0.67 | 0.68 | 0.69 | 0.69 | 0.69 | 0.69 | 0.68 | 0.67 | 0.66 |
| **Uniform** | **3** | 0.82 | 0.89 | 0.91 | 0.91 | 0.89 | 0.87 | 0.82 | 0.78 | 0.74 |
| | **5** | 0.55 | 0.59 | 0.61 | 0.61 | 0.60 | 0.58 | 0.55 | 0.52 | 0.49 |
| **Beta(.5,.5)** | **3** | 0.82 | 0.89 | 0.91 | 0.91 | 0.89 | 0.87 | 0.82 | 0.78 | 0.75 |
| | **5** | 0.55 | 0.60 | 0.61 | 0.61 | 0.60 | 0.59 | 0.55 | 0.53 | 0.50 |
| **Exponential** | **3** | 0.82 | 0.87 | 0.89 | 0.90 | 0.90 | 0.90 | 0.88 | 0.85 | 0.83 |
| | **5** | 0.62 | 0.65 | 0.66 | 0.67 | 0.67 | 0.67 | 0.65 | 0.64 | 0.62 |
| **Chi-sq(5)** | **3** | 0.82 | 0.87 | 0.90 | 0.90 | 0.90 | 0.88 | 0.85 | 0.82 | 0.79 |
| | **5** | 0.59 | 0.63 | 0.64 | 0.65 | 0.64 | 0.63 | 0.61 | 0.59 | 0.57 |
| **Weibull (shape=0.5)** | **3** | 0.83 | 0.83 | 0.84 | 0.85 | 0.87 | 0.88 | 0.90 | 0.90 | 0.91 |
| | **5** | 0.72 | 0.72 | 0.73 | 0.74 | 0.75 | 0.76 | 0.78 | 0.78 | 0.79 |
| **Weibull (shape=1.5)** | **3** | 0.82 | 0.88 | 0.90 | 0.91 | 0.90 | 0.88 | 0.85 | 0.81 | 0.78 |
| | **5** | 0.58 | 0.62 | 0.64 | 0.64 | 0.63 | 0.62 | 0.60 | 0.57 | 0.55 |
| **Pareto (shape=2.5)** | **3** | 0.83 | 0.81 | 0.80 | 0.81 | 0.82 | 0.83 | 0.84 | 0.86 | 0.87 |
| | **5** | 0.77 | 0.75 | 0.74 | 0.75 | 0.76 | 0.77 | 0.78 | 0.79 | 0.80 |
| **Pareto (shape=4)** | **3** | 0.83 | 0.84 | 0.85 | 0.86 | 0.87 | 0.87 | 0.88 | 0.88 | 0.87 |
| | **5** | 0.69 | 0.70 | 0.71 | 0.72 | 0.73 | 0.73 | 0.74 | 0.73 | 0.73 |

Table 2. RE for two $H_B$, some $H_J$ for some distributions in $n=60$

| | $H_b$ | $H_j=3$ | $H_j=4$ | $H_j=5$ | $H_j=6$ | $H_j=7$ | $H_j=8$ | $H_j=10$ | $H_j=12$ | $H_j=14$ |
|---|---|---|---|---|---|---|---|---|---|---|
| **Normal** | **3** | 0.96 | 1.16 | 1.34 | 1.50 | 1.65 | 1.78 | 1.97 | 2.05 | 2.04 |
| | **5** | 0.67 | 0.80 | 0.93 | 1.04 | 1.14 | 1.23 | 1.36 | 1.42 | 1.41 |
| **t(3)** | **3** | 0.96 | 1.06 | 1.14 | 1.19 | 1.24 | 1.27 | 1.30 | 1.31 | 1.31 |
| | **5** | 0.79 | 0.86 | 0.93 | 0.97 | 1.01 | 1.03 | 1.06 | 1.07 | 1.06 |
| **Uniform** | **3** | 0.96 | 1.18 | 1.39 | 1.58 | 1.76 | 1.91 | 2.14 | 2.24 | 2.22 |
| | **5** | 0.64 | 0.79 | 0.93 | 1.05 | 1.17 | 1.28 | 1.43 | 1.49 | 1.48 |
| **Beta(.5,.5)** | **3** | 0.96 | 1.18 | 1.37 | 1.56 | 1.73 | 1.87 | 2.09 | 2.19 | 2.17 |
| | **5** | 0.65 | 0.79 | 0.93 | 1.05 | 1.16 | 1.26 | 1.41 | 1.48 | 1.46 |
| **Exponential** | **3** | 0.96 | 1.11 | 1.24 | 1.35 | 1.45 | 1.54 | 1.66 | 1.73 | 1.75 |
| | **5** | 0.72 | 0.83 | 0.93 | 1.01 | 1.09 | 1.15 | 1.24 | 1.29 | 1.31 |
| **Chi-sq(5)** | **3** | 0.96 | 1.14 | 1.29 | 1.43 | 1.55 | 1.66 | 1.81 | 1.89 | 1.89 |
| | **5** | 0.69 | 0.81 | 0.93 | 1.03 | 1.11 | 1.19 | 1.30 | 1.35 | 1.36 |
| **Weibull (shape=0.5)** | **3** | 0.96 | 1.02 | 1.07 | 1.10 | 1.13 | 1.15 | 1.18 | 1.20 | 1.22 |
| | **5** | 0.83 | 0.89 | 0.93 | 0.96 | 0.98 | 1.00 | 1.03 | 1.04 | 1.06 |
| **Weibull (shape=1.5)** | **3** | 0.96 | 1.15 | 1.31 | 1.46 | 1.59 | 1.71 | 1.88 | 1.96 | 1.96 |
| | **5** | 0.68 | 0.81 | 0.93 | 1.03 | 1.13 | 1.21 | 1.33 | 1.38 | 1.39 |
| **Pareto (shape=2.5)** | **3** | 0.96 | 0.99 | 1.00 | 1.01 | 1.01 | 1.00 | 0.99 | 0.97 | 0.96 |
| | **5** | 0.89 | 0.91 | 0.92 | 0.93 | 0.93 | 0.93 | 0.91 | 0.90 | 0.89 |
| **Pareto (shape=4)** | **3** | 0.96 | 1.04 | 1.10 | 1.15 | 1.19 | 1.22 | 1.26 | 1.28 | 1.28 |
| | **5** | 0.81 | 0.87 | 0.93 | 0.97 | 1.00 | 1.02 | 1.05 | 1.07 | 1.08 |

Moreover, Table 2 shows that, in large sample sizes and for fixed $H_J$ and $H_B$, RE increases with $\delta_I$. But Table 1 shows that RE decreases with $\delta_I$, in small sample sizes .

Based on the above results, we propose that BRSS is preferred to JPS, in small sample sizes, while the standard JPS approach is preferable to BRSS, in large sample sizes and in situations in which we can use a larger class size for *H* in JPS set-up.

Finally, we examine the relative efficiency of $\tilde{\mu}_{JPS}$ with respect to $\hat{\mu}_{JPS}$. Figure 6 plots the relative efficiency of $\tilde{\mu}_{JPS}$ with respect to $\hat{\mu}_{JPS}$ as a function of the sample size for two ranking class sizes *H*=2 and 5 for some distributions. Figure 6 confirms that $\tilde{\mu}_{JPS}$ is more efficient than $\hat{\mu}_{JPS}$, but the relative efficiency is at most 1.10 (as mentioned by Frey and Feeman (2012)). Figure 6 shows that RE converges to 1 as sample size increases. Also, Figure 6 shows that RE decreases as $\delta_I$ increases. Therefore $\tilde{\mu}_{JPS}$ is more efficient than $\hat{\mu}_{JPS}$ only when sample size and $\delta_I$ are small.

# 4 Optimal ranking class sizes

As mentioned in the previous Subsection 3.2, $\mathrm{RE}\left(\hat{\mu}_{JPS}, \hat{\mu}_{SRS}\right)$ is not necessarily an increasing function of ranking class size *H*. In this section, we obtain the optimal *H* which minimizes the variance of the standard JPS mean estimator and thus provides maximum $\mathrm{RE}\left(\hat{\mu}_{JPS}, \hat{\mu}_{SRS}\right)$, in perfect situations. The optimal *H* ($H_{opt}$) is calculated numerically for different sample sizes from 5 to 50 (5, 10, 15, ..., 50) for some distributions and the results are presented in Table 3. Table 3 also presents the maximum of $\mathrm{RE}\left(\hat{\mu}_{JPS}, \hat{\mu}_{SRS}\right)$, *MRE*. Note that REs are calculated to two decimal places, and, in the event that more than one value of *H* correspond to the MRE, the smallest value is taken to be $H_{opt}$.

Note that, in Table 3, $H_{opt}$ being equal to 1 means that SRS mean estimator is more efficient than the standard JPS mean estimator.

Guided by Table 3, we recommend that, when population distribution is not very skewed or heavy tailed (such that $\delta_I$ is not small), $H_{opt}$ be chosen according to Table 4.

Table 3. $H_{opt}$ and *MRE*, for different sample sizes and for some distributions

| | | n=5 | n=10 | n=15 | n=20 | n=25 | n=30 | n=35 | n=40 | n=45 | n=50 |
|---|---|---|---|---|---|---|---|---|---|---|---|
| **Normal** | $H_{opt}$ | 4 | 5 | 5 | 6 | 7 | 8 | 9 | 9 | 10 | 11 |
| | $MRE_{opt}$ | 1.16 | 1.45 | 1.73 | 2.01 | 2.27 | 2.52 | 2.77 | 3.01 | 3.25 | 3.48 |
| **t(3)** | $H_{opt}$ | 5 | 5 | 5 | 5 | 6 | 7 | 8 | 9 | 10 | 10 |
| | $MRE_{opt}$ | 1.06 | 1.18 | 1.29 | 1.40 | 1.49 | 1.58 | 1.66 | 1.74 | 1.81 | 1.87 |
| **Uniform** | $H_{opt}$ | 4 | 4 | 5 | 6 | 7 | 8 | 9 | 9 | 10 | 11 |
| | $MRE_{opt}$ | 1.18 | 1.50 | 1.83 | 2.15 | 2.46 | 2.76 | 3.06 | 3.35 | 3.64 | 3.93 |
| **Beta(.5,.5)** | $H_{opt}$ | 4 | 5 | 5 | 6 | 7 | 8 | 9 | 9 | 10 | 11 |
| | $MRE_{opt}$ | 1.17 | 1.49 | 1.80 | 2.10 | 2.40 | 2.69 | 2.97 | 3.24 | 3.52 | 3.79 |
| **Exponential** | $H_{opt}$ | 5 | 5 | 7 | 7 | 8 | 8 | 9 | 10 | 11 | 12 |
| | $MRE_{opt}$ | 1.10 | 1.29 | 1.48 | 1.65 | 1.82 | 1.99 | 2.14 | 2.29 | 2.44 | 2.58 |
| **Chi-sq(5)** | $H_{opt}$ | 4 | 5 | 6 | 6 | 7 | 8 | 9 | 10 | 11 | 11 |
| | $MRE_{opt}$ | 1.13 | 1.37 | 1.61 | 1.83 | 2.04 | 2.25 | 2.46 | 2.65 | 2.84 | 3.02 |
| **Weibull** | $H_{opt}$ | 14 | 14 | 16 | 13 | 14 | 16 | 15 | 18 | 17 | 16 |
| **(shape=0.5)** | $MRE_{opt}$ | 1.03 | 1.08 | 1.13 | 1.17 | 1.22 | 1.27 | 1.31 | 1.36 | 1.40 | 1.44 |
| **Weibull** | $H_{opt}$ | 4 | 5 | 6 | 6 | 7 | 8 | 9 | 10 | 11 | 11 |
| **(shape=1.5)** | $MRE_{opt}$ | 1.14 | 1.40 | 1.65 | 1.89 | 2.12 | 2.35 | 2.57 | 2.78 | 2.99 | 3.19 |
| **Pareto** | $H_{opt}$ | **1** | 20 | 23 | 24 | 2 | 3 | 4 | 4 | 4 | 5 |
| **(shape=2.5)** | $MRE_{opt}$ | 1.00 | 1.01 | 1.02 | 1.03 | 1.04 | 1.06 | 1.08 | 1.10 | 1.11 | 1.13 |
| **Pareto** | $H_{opt}$ | 7 | 9 | 9 | 9 | 10 | 9 | 11 | 11 | 12 | 11 |
| **(shape=4)** | $MRE_{opt}$ | 1.04 | 1.12 | 1.19 | 1.26 | 1.33 | 1.39 | 1.46 | 1.52 | 1.58 | 1.63 |

Table 4. Recommended $H_{opt}$ for different sample size

| | n=5 | n=10 | n=15 | n=20 | n=25 | n=30 | n=35 | n=40 | n=45 | n=50 |
|---|---|---|---|---|---|---|---|---|---|---|
| $H_{opt}$ | 4 | 5 | 6 | 6 | 7 | 8 | 9 | 10 | 10 | 11 |

# 5 Conclusion

In this paper, we discussed some of the properties of non-parametric estimation of expectation of *g(X)* (any function of *X*) by using a JPS sample. We considered a class of estimators and provided its mean and variance. We showed that an estimator in the class is an unbiased and under some condition is consistent, and asymptotically normally distributed. We compared the performance of an estimator in the class with respect to the corresponding SRS and BRSS estimators. We showed that any estimator in the class may be less efficient than SRS estimator for small sample sizes. We proved that the relative efficiency of any estimator in the class with respect to BRSS estimator tends to 1 as the sample size goes to infinity. As examples of *g(X)*, we illustrated the results for the estimation of population mean, population variance and CDF. We specifically examined the estimation of population mean. We showed that the standard JPS mean estimator may be more efficient than BRSS for large sample sizes, since JPS is more flexible than BRSS for choosing class size *H*, in some applications. We saw that, although the JPS mean estimator is dominated by the optimum JPS mean estimator, proposed by Frey and Feeman (2012), the reduction in variance is at most 10% which occurs in small sample sizes and for some distributions. However as sample size increases the relative efficiency of one with respect to the other goes to one and the two estimators have the same asymptotic distribution. In addition, we showed that ranking class size, $H$, should not be taken anything large, even from a theoretical point of view, because of the probability that the JPS sample has empty strata, increases with $H$, thus the variance of the standard JPS mean estimator increases when $H$ becomes too large. We obtained the optimal $H$ such that it provides the minimum variance of the standard JPS mean estimator for different sample sizes, for some distribution, and recommend the way $H$ should be chosen non-parametrically for different sample sizes when population distribution is not very skewed or heavy tailed.

# Figures

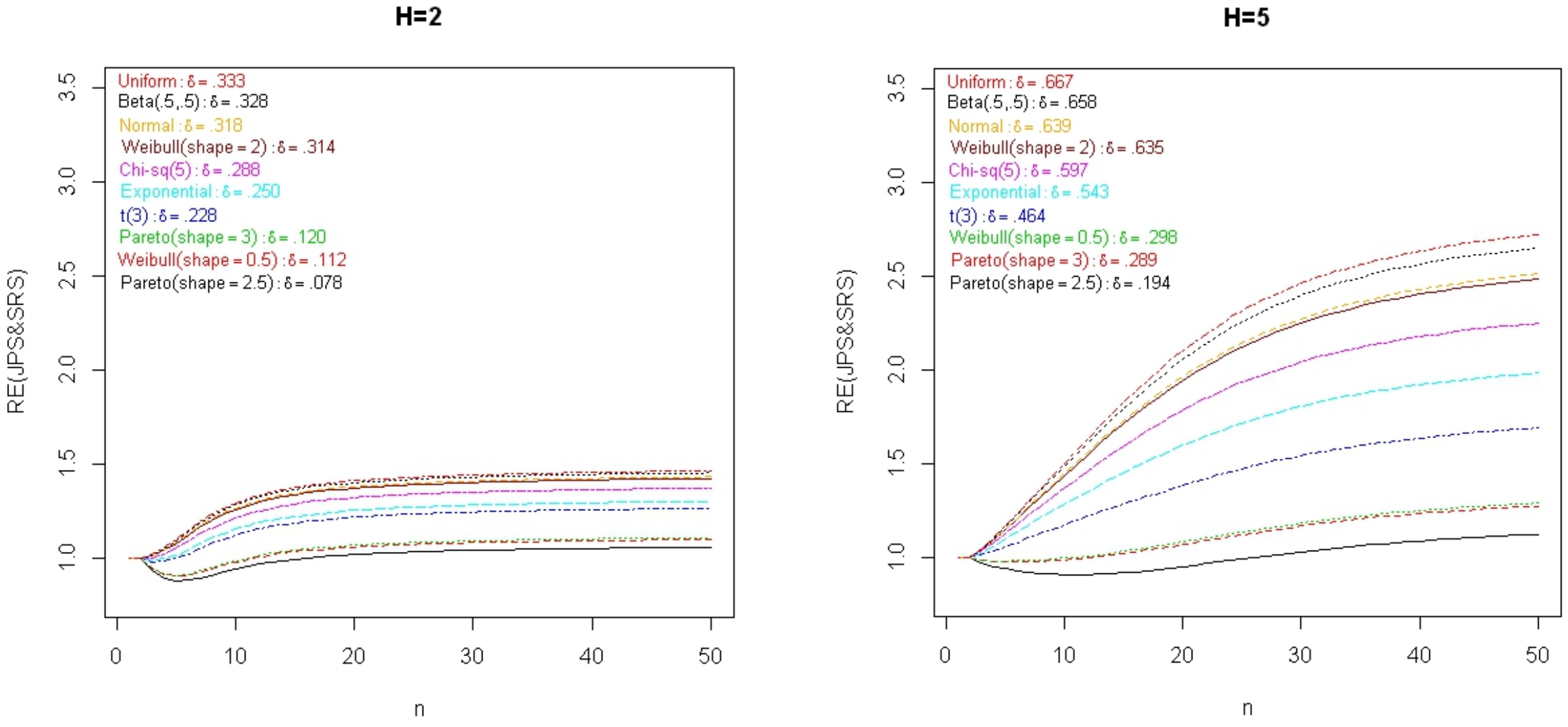

(a) (b)

Figure1: RE of standard JPS mean estimator with respect to SRS mean estimator as a function of *n* for *H*=2 and 5 for some distributions.

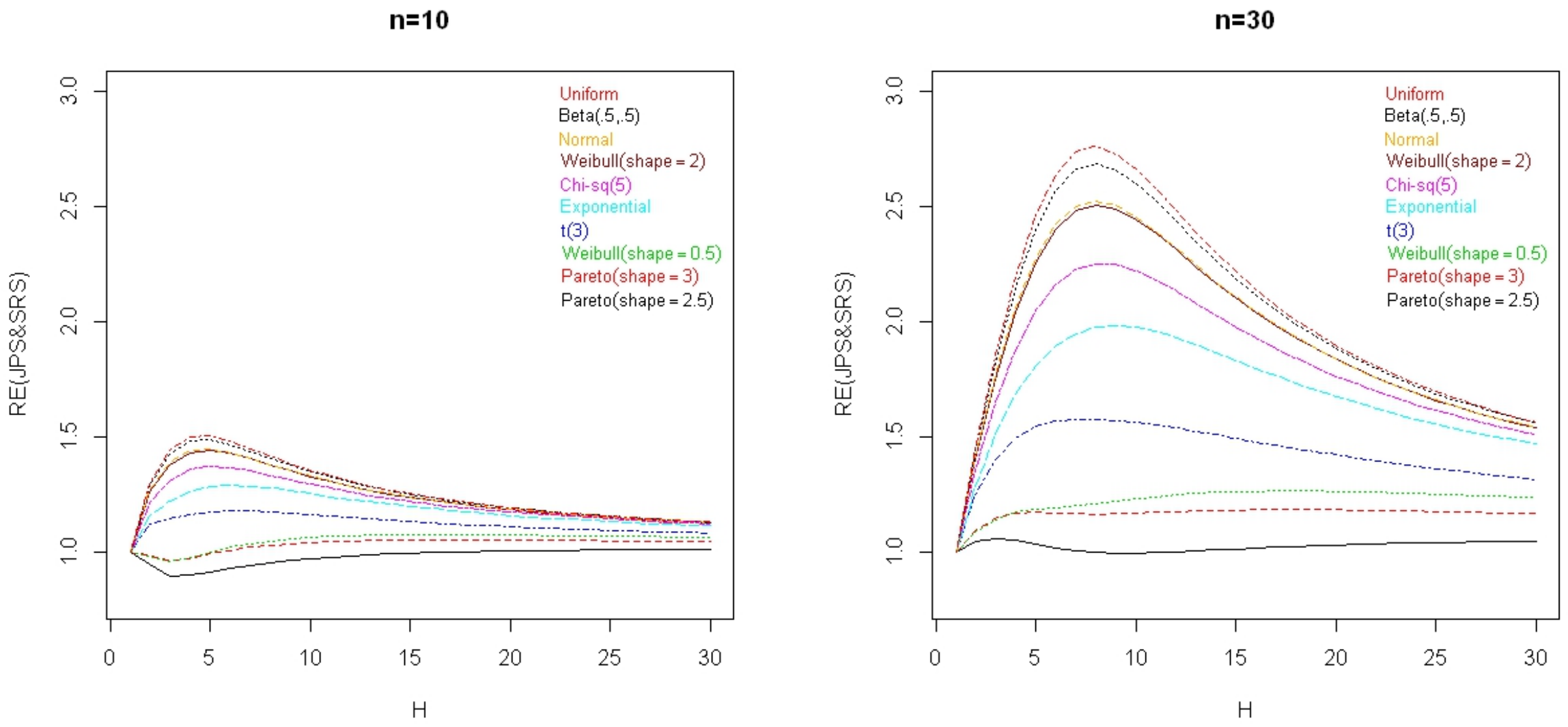

(a) (b)

Figure 2: RE of standard JPS mean estimator with respect to SRS mean estimator as a function of *H* for *n* =10 and 30 and some distributions.

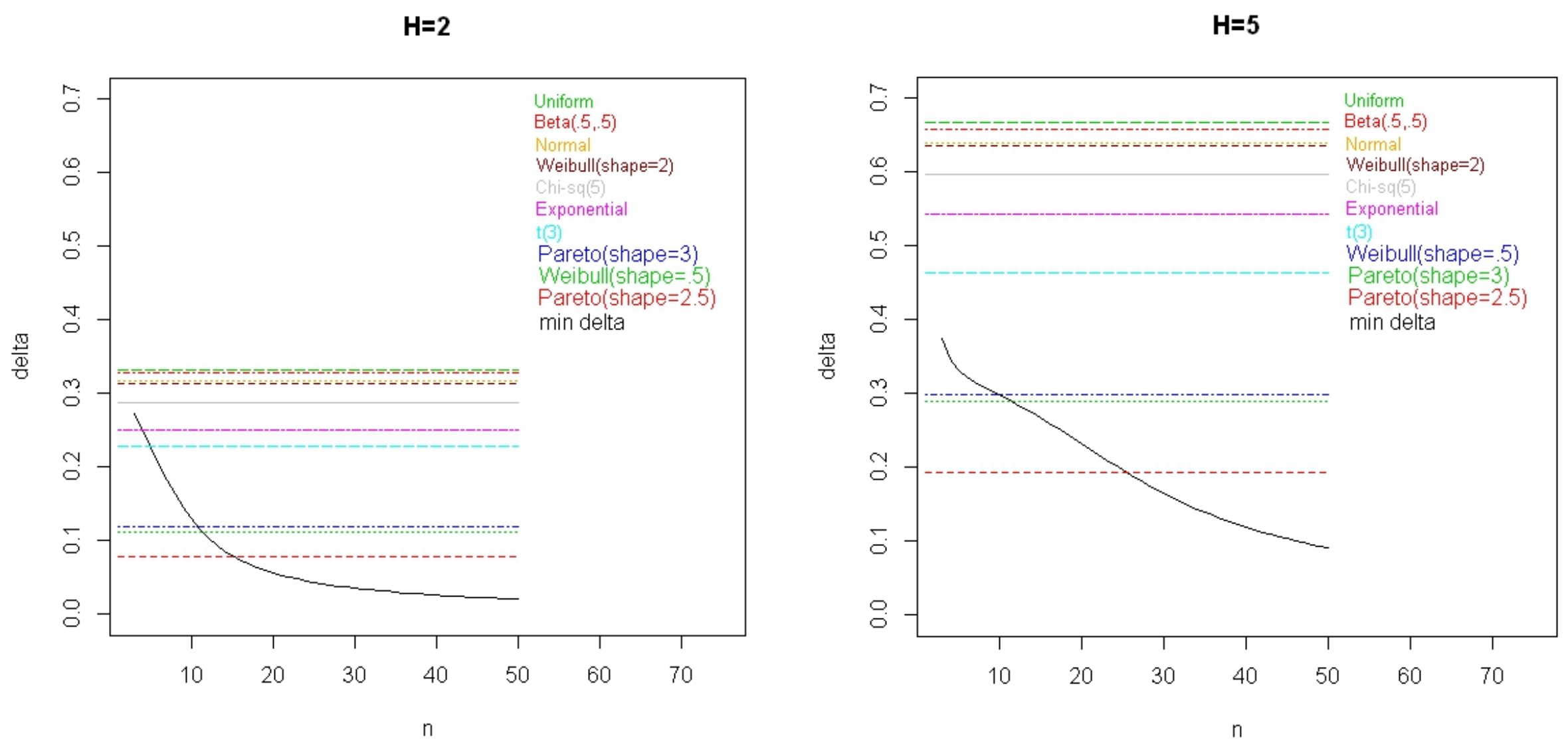

(a) (b)

Figure 3: Minimum value of $\delta$ for which RE≥1 as a function of $n$ for H=2 and 5.

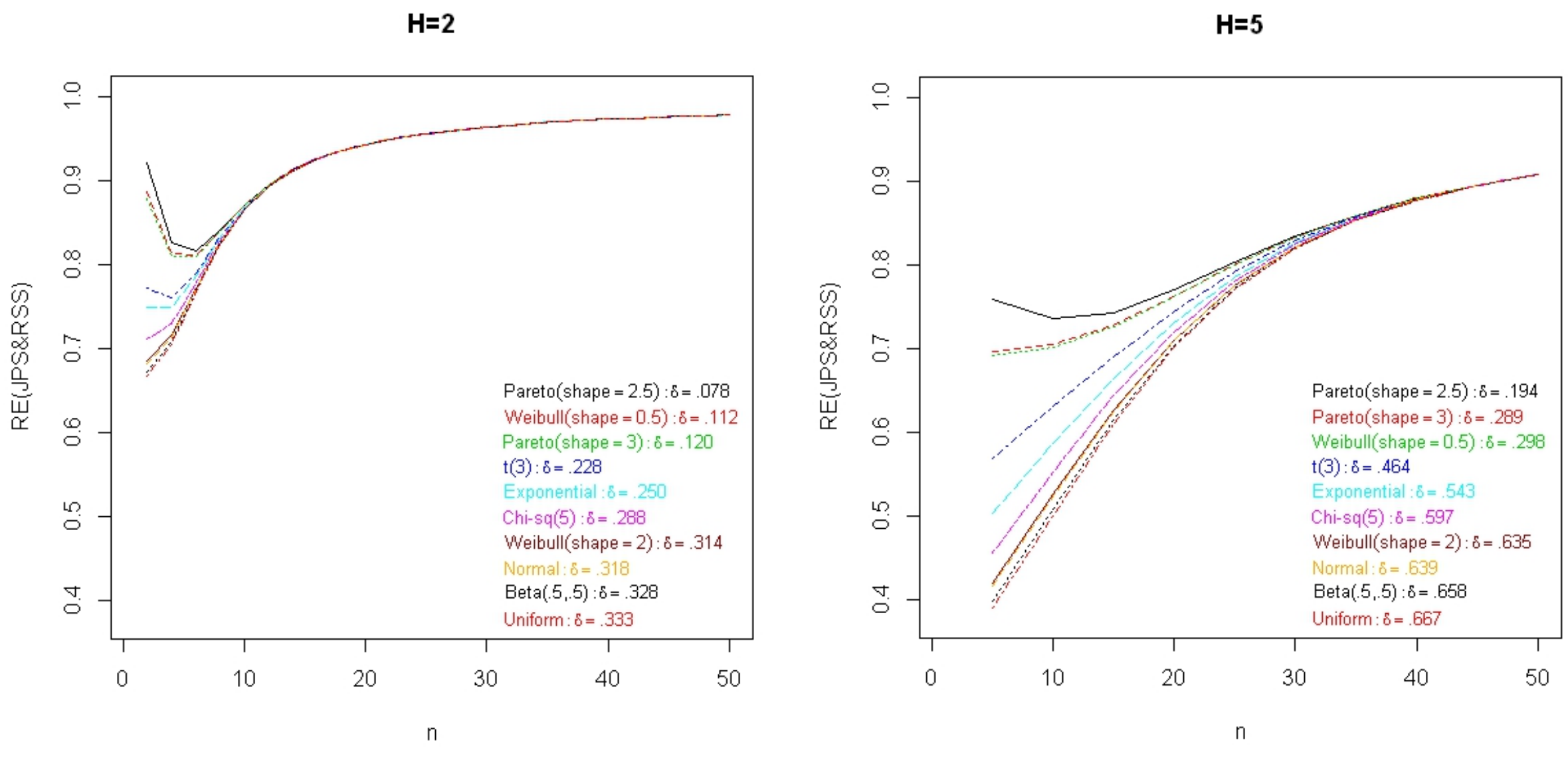

(a) (b)

Figure4: RE of standard JPS mean estimator with respect to BRSS mean estimator as a function of $n$ for $H$=2 and 5 for some distributions

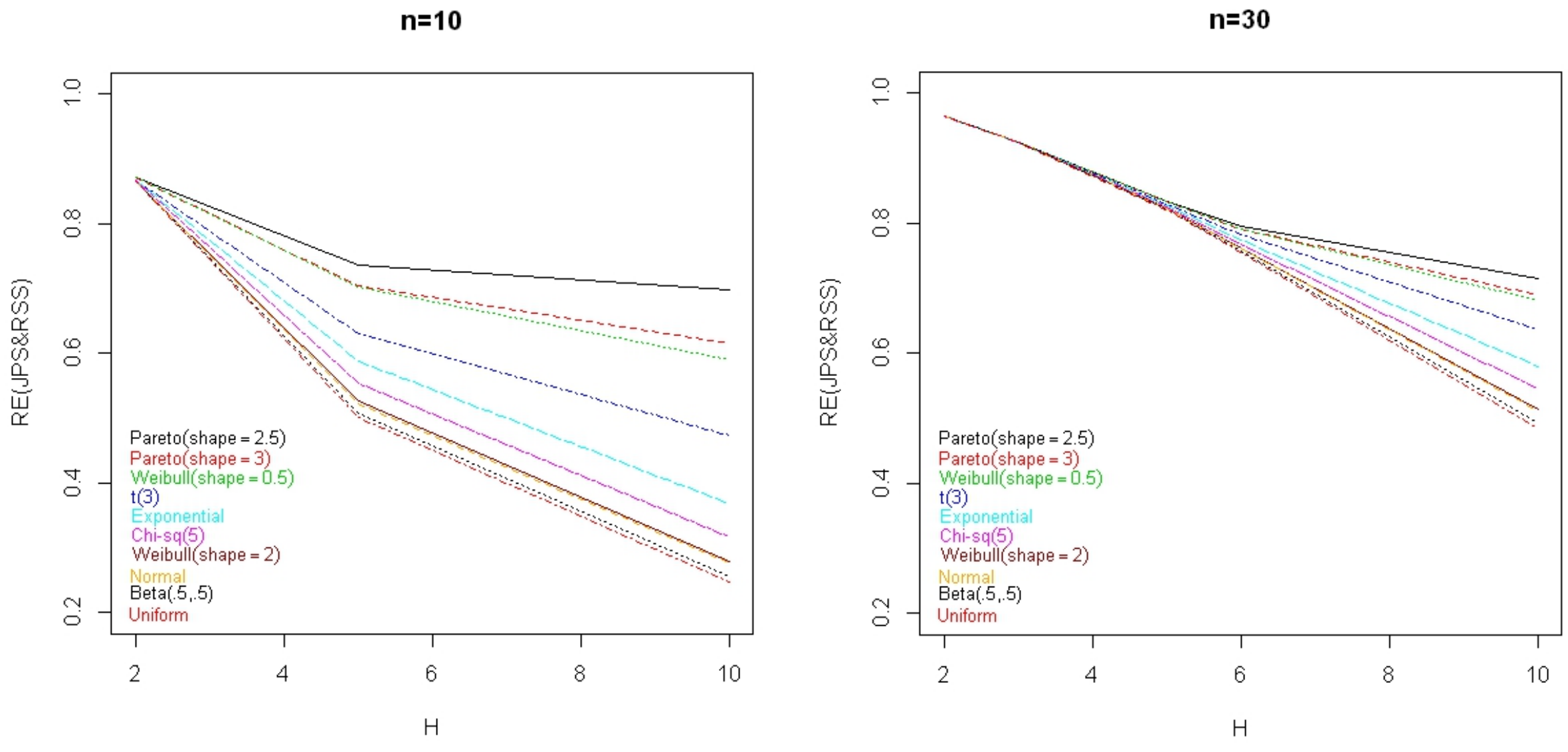

(a) (b)

Figure 5: RE of standard JPS mean estimator with respect to BRSS mean estimator as a function $H$ of for $n$ =10 and 30 and for some distributions

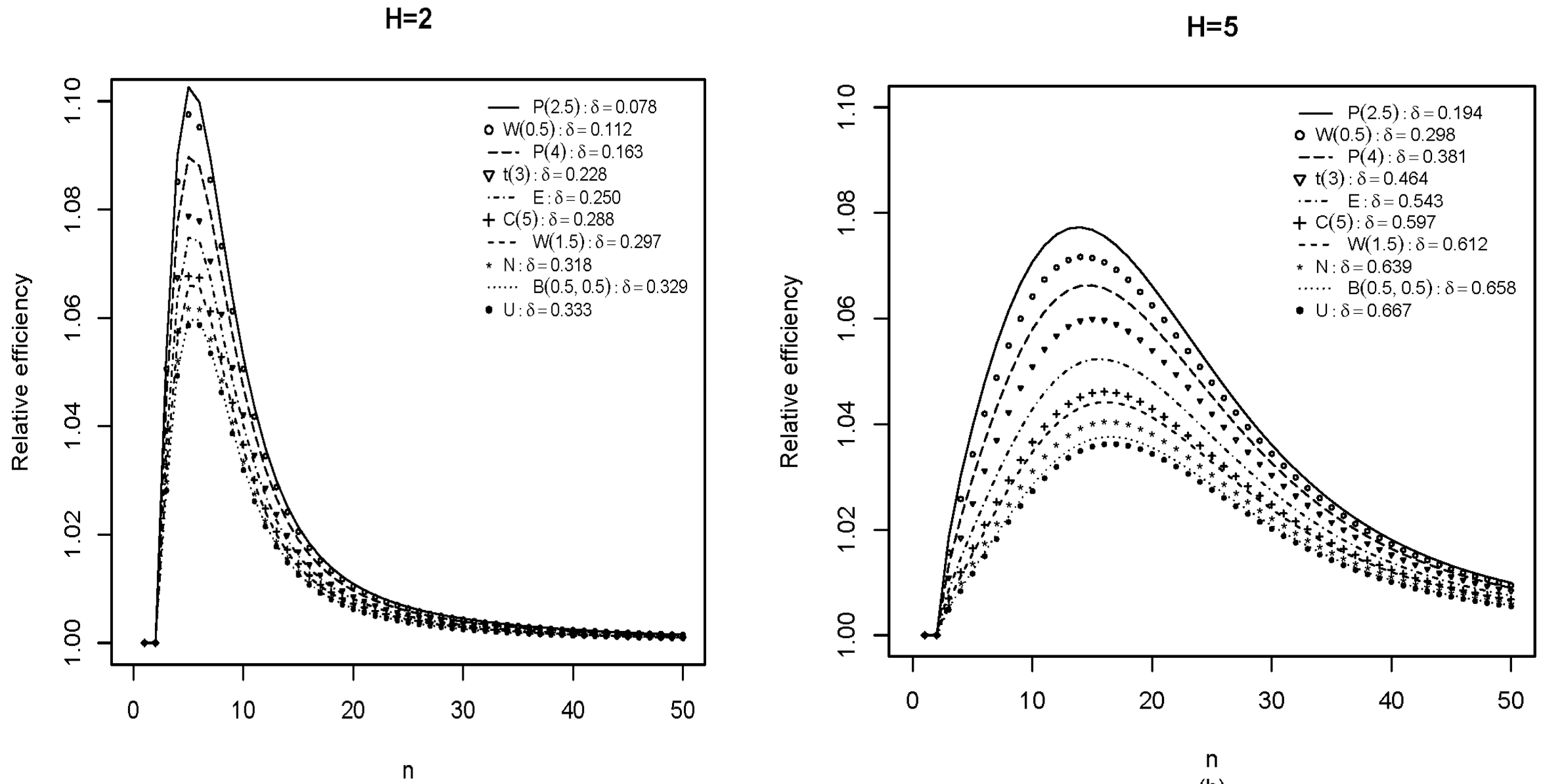

(a) (b)

Figure6: RE of Frey & Feeman JPS mean estimator with respect to standard JPS mean estimator as a function of $n$ for $H$=2 and 5 for some distributions

# Appendix

**A.**

**Lemma 3.** (i) $E\left(C_r\right)=\frac{1}{H}\quad \forall r$;

(ii) $COV\left(C_r,C_s\right)=\frac{-1}{H-1}\mathrm{V}\left(C_r\right)\ \forall r\neq s$;

(iii) For any r, $nH\,E\left(J_rC_r^2\right)\geq 1$ and the equality holds if and only if $C_r=\frac{N_r}{n}$.

***Proof:*** Note that $C_r$ 's are identity distributed.

(i) Let $A=E\left(C_1\right)=...=E\left(C_H\right)$. So, $H\times A=E\left(\sum_{r=1}^{H}C_r\right)=1$, which results in $A=\frac{1}{H}$.

(ii) First, note that $V\left(\sum_{r=1}^{H}C_r\right)=0$. So

$$\sum_r V\left(C_r\right)+2\sum_{r<s}COV\left(C_r,C_s\right)=HV\left(C_1\right)+H\left(H-1\right)COV\left(C_1,C_2\right)=0.$$

And this gives the result.

(iii) We have $nH\,E\left(J_rC_r^2\right)=H^2E\left(N_r\right)E\left(J_rC_r^2\right)\geq H^2\left[E\left(C_r\right)\right]^2=1$. The inequality holds by the Cauchy-Schwarz inequality and the equality holds if and only if there exist real constants $a$ and $b$ such that $\sqrt{J_r}C_r\overset{a.s.}{=}a\sqrt{N_r}+b$ or equivalently $C_r\overset{a.s.}{=}aN_r+b\sqrt{N_r}$. Hence, $1\overset{a.s.}{=}a\times n+b\sum_{r=1}^{H}\sqrt{N_r}$. It must be $b=0$. So, $C_r\overset{a.s.}{=}\frac{N_r}{n}$. □

**B.**

The following Lemma gives the distribution of $\frac{I_r}{h_n}$, $r=1,...,H$, and some of its properties.

***Lemma 4.*** *(i) Probability function of* $\frac{I_r}{h_n}$, $r=1,...,H$, *is given by*

$$P\left(\frac{I_r}{h_n}=u\right)=\begin{cases}\left(\frac{H-1}{H}\right)^n & u=0\\ \frac{1}{H^n}\binom{H-1}{k-1}\sum_{j=1}^{k}\left(-1\right)^{j-1}\binom{k}{j-1}\left(k-j+1\right)^n & u=\frac{1}{k};\ k=1,...,H\end{cases}$$

$(ii)\ E\left(\frac{I_r}{h_n}\right)=\frac{1}{H};$

$(iii)\ V\left(\frac{I_r}{h_n}\right)=\frac{1}{H^2}\sum_{k=1}^{H-1}\left(\frac{k}{H}\right)^{n-1};$

$(iv)\ COV\left(\frac{I_r}{h_n},\frac{I_s}{h_n}\right)=\frac{-1}{H-1}\mathrm{V}\left(\frac{I_r}{h_n}\right)\ \forall r\neq s;$

$(v)\ E\left(\frac{J_r}{h_n^2}\right)=\frac{1}{H^n}\left[\frac{1}{n}+\sum_{h_n=2}^{H}\sum_{j=1}^{h_n-1}\sum_{n_1=1}^{n-h_n+1}\frac{(-1)^{j-1}}{h_n^2 n_1}\binom{H-1}{h_n-1}\binom{h_n-1}{j-1}\binom{n}{n_1}(h_n-j)^{n-n_1}\right];$

$(vi)\ \frac{nH^2}{H-1}V\left(\frac{I_1}{h_n}\right)<1\ \ \forall n\geq 3;$

$(vii)\ \sqrt{n}\left(\frac{1}{h_n}-\frac{1}{H}\right)\xrightarrow{a.s}0;$

$(viii)\ nV\left(\frac{I_1}{h_n}\right)\xrightarrow{n\to\infty}0.$

***Proof***: The parts (ii) and (iv) holds because of the parts (i) and (ii) of Lemma 3, respectively. Note that $\frac{I_r}{h_n}$'s are identically distributed. Hence, without loss totality, we can consider $\frac{I_1}{h_n}$.

(i) $P\left(\frac{I_1}{h_n}=0\right)=P\left(N_1=0\right)=\left(\frac{H-1}{H}\right)^n$ and, for each $k=1,\dots,H$,

$$
\begin{aligned}
P\left(\frac{I_1}{h_n}=\frac{1}{k}\right)&=P\left(N_1>0,h_n=k\right)\\
&=\binom{H-1}{k-1}P\left(N_1>0,\dots,N_k>0,\ N_i=0;i>k\right)\\
&=\binom{H-1}{k-1}\sum_{n_1,\dots,n_k>0}\binom{n}{n_1,\dots,n_k}\frac{1}{H^n}\\
&=\frac{1}{H^n}\binom{H-1}{k-1}\sum_{j=1}^{k}(-1)^{j-1}\binom{k}{j-1}(k-j+1)^n.
\end{aligned}
$$

The last equality holds by induction and the equality

$$\sum_{n_1,\dots,n_H=n}\binom{n}{n_1,\dots,n_H}=\sum_{k=1}^{H}\binom{H}{k}\sum_{n_1,\dots,n_k>0}\binom{n}{n_1,\dots,n_k}.$$

(iii)

$$E\left(\frac{I_1}{h_n^2}\right)=\sum_{h_n=1}^{H}\frac{\binom{H-1}{h_n-1}}{h_n^2H^n}\sum_{n_1,\dots,n_{h_n}>0}\binom{n}{n_1,\dots,n_{h_n}}$$

$$=\sum_{h_n=1}^{H}\frac{\binom{H-1}{h_n-1}}{h_n^2H^n}\sum_{j=1}^{h_n}(-1)^{j-1}\binom{h_n}{j-1}(h_n-j+1)^n$$

$$=\frac{1}{H^n}\sum_{k=1}^{H}\left(\sum_{j=1}^{H-k+1}\frac{(-1)^{j-1}}{k+j-1}\binom{H-1}{j-1}\binom{H-j}{k-1}\right)k^{n-1}$$

$$=\frac{1}{H^2}\sum_{k=1}^{H}\left(\frac{k}{H}\right)^{n-1}.$$

(v)

$$E\left(\frac{J_1}{h_n^2}\right)=\sum_{h_n=1}^{H}\frac{\binom{H-1}{h_n-1}}{h_n^2H^n}\sum_{n_1,\dots,n_{h_n}>0}\frac{1}{n_1}\binom{n}{n_1,\dots,n_{h_n}}$$

$$=\frac{1}{H^n}\left[\frac{1}{n}+\sum_{h_n=2}^{H}\frac{\binom{H-1}{h_n-1}}{h_n^2}\sum_{n_1=1}^{n-h_n+1}\frac{1}{n_1}\binom{n}{n_1}\sum_{n_2,\dots,n_{h_n}>0}\binom{n-n_1}{n_2,\dots,n_{h_n}}\right]$$

$$=\frac{1}{H^n}\left[\frac{1}{n}+\sum_{h_n=2}^{H}\sum_{j=1}^{h_n-1}\sum_{n_1=1}^{n-h_n+1}\frac{(-1)^{j-1}}{h_n^2n_1}\binom{H-1}{h_n-1}\binom{h_n-1}{j-1}\binom{n}{n_1}(h_n-j)^{n-n_1}\right].$$

(vi) Using (iii), we have

$$\frac{nH^2}{H-1}V\left(\frac{I_1}{h_n}\right)=\frac{n}{H-1}\sum_{k=1}^{H-1}\left(\frac{k}{H}\right)^{n-1}.$$

Let $f(x)=\frac{1}{H-1}\sum_{k=1}^{H-1}x\left(\frac{k}{H}\right)^{x-1}, x>1$. Hence,

$$\frac{df(x)}{dx}=\frac{1}{H-1}\sum_{k=1}^{H-1}\left[\left(\frac{k}{H}\right)^{x-1}\left(1+x\ln\left(\frac{k}{H}\right)\right)\right]<0.$$

So, $f(x)$ is a decreasing function. Besides $f(1)=f(2)=1$. Thus for fixed $H$ and any $n \geq 3$

$\frac{n}{H-1}\sum_{k=1}^{H-1}\left(\frac{k}{H}\right)^{n-1} < 1$.

(vii) We know $1/h_n$ converges a.s. to $1/H$. So

$$\exists F \text{ s.t. } P(F)=1 \;\&\; \forall \omega \in F \;\; \frac{1}{h_n(\omega)} \xrightarrow{n\to\infty} \frac{1}{H};$$

that is, when $\omega \in F$,

$$\forall \varepsilon \; \exists n_0(\omega) \text{ s.t. } \forall n > n_0(\omega) \;\; 0 \leq \frac{1}{h_n(\omega)} - \frac{1}{H} < \varepsilon.$$

We fix $\omega$ and set $\varepsilon = \frac{1}{H-1} - \frac{1}{H}$. Hence

$$\exists n_1(\omega) \; s.t. \; \forall n > n_1(\omega) \;\; \frac{1}{h_n(\omega)} - \frac{1}{H} = 0.$$

Thus

$$\forall \varepsilon \leq \frac{1}{H-1} - \frac{1}{H} \;\&\; n > n_1(\omega), \;\; \sqrt{n}\left(\frac{1}{h_n(\omega)} - \frac{1}{H}\right) = 0.$$

And this completes the proof.

(viii) We have $nV(I_1/h_n) = E\left[\sqrt{n}\left(I_1/h_n - 1/H\right)\right]^2$. And part (vii) completes the proof.□

**C.**

***Lemma 5.*** $(i)\; E(A_r) = \frac{1}{H} \;\; \forall r;$

$(ii)\; COV(A_r, A_s) = \frac{-1}{H-1} \mathrm{V}(A_r) \;\; \forall r \neq s;$

$(iii)\; \frac{nH^2}{H-1} V(A_r) < 1 \;\; \forall n \geq 3;$

$(iv)\; nH\, E(J_1 A_1^2) \xrightarrow{n\to\infty} 1;$

$(v)\; \sqrt{n}\left(A_r - \frac{1}{H}\right) \xrightarrow{a.s} 0 \;\; \forall r;$

$(vi)\ nV\left(A_r\right)\xrightarrow{n\to\infty}0\ \forall r.$

***Proof:*** The parts (i) and (ii) holds because of the parts (i) and (ii) of Lemma 3, respectively.

(iii) We have

$$V\left(\hat{\mu}_{JPS}\right)-V\left(\tilde{\mu}_{JPS}\right)=\left[E\left(\frac{J_1}{h_n^2}\right)-E\left(J_1A_1^2\right)\right]\sum_{r=1}^{H}\sigma_g^2[r]$$

$$+\frac{H}{H-1}\left[V\left(\frac{I_1}{h_n}\right)-V\left(A_1\right)\right]\sum_{r=1}^{H}\left(\mu_g[r]-\mu_g\right)^2.$$

Since $\tilde{\mu}_{JPS}$ dominates $\hat{\mu}_{JPS}$,

$$\frac{nH^2}{H-1}V\left(A_r\right)\le\frac{nH^2}{H-1}V\left(\frac{I_r}{h_n}\right)<1\ \ \forall n\ge 3.$$

(iv) For any $r$, $\frac{N_r}{n}$ converges a.s. to $\frac{1}{H}$ as $n\to\infty$. Hence, $A_1$ converges a.s. to $\frac{1}{H}$ as $n\to\infty$. Also, $nJ_1$ converges a.s. to $H$ as $n\to\infty$. Thus, $nHJ_1A_1^2$ converges a.s. to 1 as $n\to\infty$. And this gives the result.

(v) We saw that $A_1$ converges a.s. to $\frac{1}{H}$ as $n\to\infty$. That is

$$\exists F \text{ s.t. } P\left(F\right)=1\ \&\ \ \forall\omega\in F\ \ \&\ \varepsilon>0\ \exists n_0\left(\omega\right)\ s.t.\ \forall n>n_0\left(\omega\right)\ \left|A_1\left(\omega\right)-\frac{1}{H}\right|<\varepsilon.$$

Hence for fix $\omega$

$$\forall\varepsilon\le\min_{L>0}\left(L\right)\ \ \exists n_1\left(\omega\right)\ s.t.\ \forall n>n_1\left(\omega\right)\ \ \ L=0,$$

where $L=\left|A_1\left(\omega\right)-\frac{1}{H}\right|$. So,

$$\forall\varepsilon\le\min_{L>0}\left(L\right)\ \ \exists n_1\left(\omega\right)\ s.t.\ \forall n>n_1\left(\omega\right)\ \ \ \sqrt{n}L=0.$$

That is $\sqrt{n}\left(A_1-\frac{1}{H}\right)$ converges a.s. to 0 as $n\to\infty$.

(vi) We have $nV\left(A_1\right)=E\left[\sqrt{n}\left(A_1-1/H\right)\right]^2$. And part (v) completes the proof.□

**D.**

***Lemma 6.*** *Let* $A = \{a_1 \le a_2 \le ... \le a_k\}$ *and* $B = \{b_1 \le b_2 \le ... \le b_k\}$ *be two sets of real numbers. Then*

$$\sum_{j=1}^{k} a_j b_{l_j} \tag{5}$$

*is maximized when* $l_1 l_2 ... l_k = 12...k =$ *the natural permutation.*

***Proof:*** We first note that without loss of generality (w.l.g) we can assume $a_1 = b_1 = 0$. With that assumption, (5) is obvious for $k = 2$.

Now assume (5) holds for $k = r \ge 2$. We claim that it holds for $k = r + 1$ and thus the lemma is proved by induction.

To prove the claim, again assume (w.l.g) that $a_1 = b_1 = 0$. It is, then, obvious that $l_1 = 1$ is a necessary condition for $\sum_{j=1}^{r+1} a_j b_{l_j}$ to be maximized. Thus, we can write

$$\begin{aligned} \sum_{j=1}^{r+1} a_j b_{l_j} &\le a_1 b_1 + \sum_{j=2}^{r+1} a_j b_{l_j} \\ &\le a_1 b_1 + \sum_{j=2}^{r+1} a_j b_j \\ &= \sum_{j=1}^{r+1} a_j b_j , \end{aligned}$$

where $l_2 ... l_k$ is a permutation of $2...k$ .

The second inequality follows from the induction assumption. Thus, the lemma is proved.□

***Lemma 7.*** *Let* $A$ *and* $B$ *be as in Lemma 6. Then*

$$\sum_{j=1}^{k} \left(a_j + b_j\right)^2 \ge \sum_{j=1}^{k} \left(a_j + b_{l_j}\right)^2$$

*for any permutation* $l_1 l_2 ... l_k$ *of* $12...k$ .

***Proof:*** It follows easily from Lemma 6.□

***Lemma 8.*** *Let* $A_i = \{a_{i1}, ..., a_{ik}\}$, $i = 1, ..., n$, *be n sets of real numbers. Then*

$$\sum_{j=1}^{k} \left( \sum_{i=1}^{n} a_{i(j)} \right)^2 \ge \sum_{j=1}^{k} \left( \sum_{i=1}^{n} a_{ij} \right)^2 ,$$

*where* $a_{i(j)}$ *is the* $j^{th}$ *element of ordered* $A_i$ .

***Proof:*** For $n = 2$, the truth of the lemma follows from Lemma 7. Now, suppose the lemma is true for $n = r \geq 2$. We claim that it is, then, true for $n = r + 1$ and thus the lemma is proved by induction.

To prove the claim, let

$$A = \left\{ \sum_{i=1}^{r} a_{i(1)} \leq \sum_{i=1}^{r} a_{i(2)} \leq \ldots \leq \sum_{i=1}^{r} a_{i(k)} \right\}$$

and

$$B = \left\{ a_{r+1(1)} \leq a_{r+1(2)} \leq \ldots \leq a_{r+1(k)} \right\}.$$

Then the claim is proved by Lemma 7.□